\documentclass[submission,copyright,creativecommons]{eptcs}
\providecommand{\event}{FESCA 2013} 
\usepackage{breakurl}             
\usepackage{graphicx}
\usepackage{subfigure}
\usepackage{tabularx,colortbl}
\usepackage{url}
\usepackage{multirow}

\title{Software model refactoring based on performance analysis:\\
better working on software or performance side?}

\author{
        Davide Arcelli, Vittorio Cortellessa
       \institute{Dipartimento di Ingegneria e Scienze dell'Informazione, e Matematica}
       \institute{Universit\`a degli Studi dell'Aquila}
       \email{\{davide.arcelli, vittorio.cortellessa\}@univaq.it}
}

\def\titlerunning{Software model refactoring based on performance analysis}
\def\authorrunning{D. Arcelli, V. Cortellessa}
\begin{document}
\maketitle

\begin{abstract}

Several approaches have been introduced in the last few years to tackle the
problem of interpreting model-based performance analysis results and translating
them into architectural feedback. Typically the interpretation can take
place by browsing either the software model or the performance model.
In this paper, we compare two approaches that we have recently introduced for
this goal: one based on the detection and solution of performance antipatterns,
and another one based on bidirectional model transformations between software
and performance models.
We apply both approaches to the same example in order to illustrate
the differences in the obtained performance results. Thereafter, we raise the level
of abstraction and we discuss the pros and cons of working on the software side and
on the performance side.

\end{abstract}

\section{Introduction}
\label{sec:introduction}

Identifying and removing the causes of poor performance in software systems are complex
problems due to a variety of factors to take into account. Similarly to other non-functional
properties, performance is an emergent attribute of software, as it is the result
of interactions among software components, underlying platforms, users and contexts
\cite{DBLP:conf/icse/WoodsideFP07}. The current approaches to these problems are mostly
based on the skills and experience of software developers or, in the best cases,
of performance analysts.

Quite sophisticated profiling tools have been introduced for run-time performance monitoring
\cite{perfMonitoring}, but it is well-known that the costs of solving
performance problems at runtime is orders of magnitude larger than the ones at early phases
of the software lifecycle \cite{a-Harreld98}.
Therefore, instruments that help to identify and remove causes of software performance
problems early in the lifecycle are very beneficial.

\begin{figure}[!h]
\centering
\vspace{-.1cm}
\includegraphics[width=16cm]{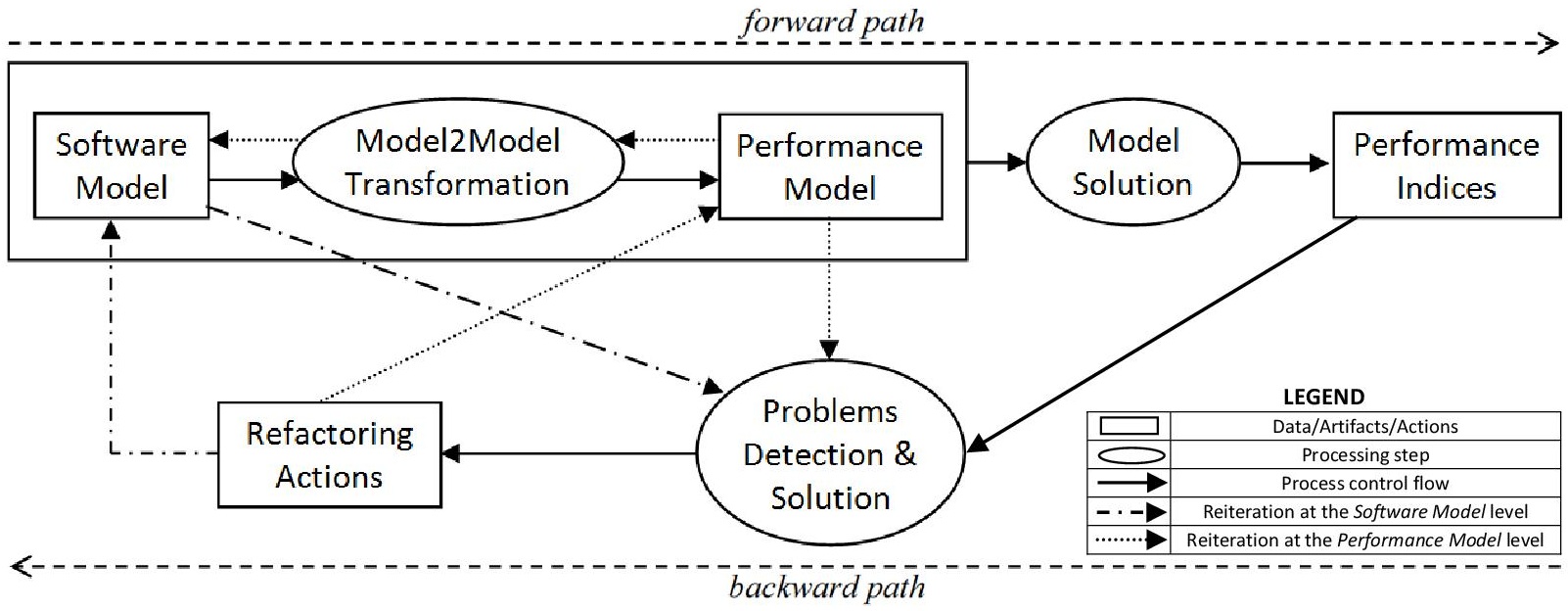}
\vspace{-.8cm}
\caption{Round-trip Software Performance Engineering.}
\label{fig:process}
\vspace{-.6cm}
\end{figure}

In Figure \ref{fig:process} a round-trip Software Performance Engineering (SPE) process
is schematically represented.
The forward path starts from a software model that is transformed
into a performance model (e.g. \cite{DBLP:conf/wosp/WoodsidePPSIM05}) that can
be solved with common performance analysis techniques/tools to obtain performance indices
\cite{Lazowska:1984:QSP:2971,DBLP:conf/wosp/CasaleS11}.
The backward path consists of a problems detection/solution step that processes the
performance indices, in conjunction with the software artifact and/or the
performance model, to detect and remove possible sources of performance problems.
Hence, a set of refactoring actions that may apply to the software artifact and/or the
performance model is obtained.
This round-trip process is reiterated until satisfactory performance indices
are obtained.

Two options have been represented in Figure \ref{fig:process} for what
concerns the reiteration mechanism, and they are identified by non-continuous arrows.
Dotted arrows represent the option of working on the performance model to detect and
remove performance problems. In this case a transformation from the performance model
to the software model has to take place when satisfactory performance indices are obtained.
Dotted-line arrows represent the option of working on the software model where refactoring
actions are applied. In both cases the forward path has to be run at each iteration
to obtain the performance indices of a refactored (performance or software) model.

In the last 5 years several approaches have appeared for identification and removal of
performance problems either in the software model \cite{DBLP:conf/iceccs/CortellessaMT10,
QoSA2012, Martens2010} or in the performance model \cite{DBLP:conf/qosa/EramoCPT12, DBLP:conf/wosp/Xu08}.
Although these two categories of approaches nicely fit into the round-trip process of
Figure \ref{fig:process}, they work in different modeling environments, under different
assumptions. Hence, not only they might achieve different results, but also they can
show very different characteristics in terms of automation, scalability, effectiveness, etc.

The goal of this paper is to highlight the differences between these two categories of
approaches in order to envisage the contexts where they can be more appropriately used.
For this goal we consider two approaches that we have recently introduced.

The first approach is based on the detection and solution, on the software model, of
\emph{performance antipatterns} that are used for ``codifying'' the knowledge and
experience of analysts by means of the identification of a \emph{problem}, i.e. a bad practice
that negatively affects software performance, and a \emph{solution}, i.e. a set of refactoring
actions that can be carried out to remove it \cite{QoSA2012}.
This approach involves dotted-line arrows of Figure \ref{fig:process}.

The second approach is based on \emph{bidirectional model transformations} between UML
software models and Queueing Network (QN) performance models.
A forward transformation is used to generate the performance model
from an initial software model. The corresponding backward transformation is
used to generate a new software model from a satisfactory performance model
obtained by means of changes made by the analyst on the performance side
\cite{DBLP:conf/qosa/EramoCPT12}.
This approach involves dotted arrows of Figure \ref{fig:process}.

We apply both the approaches to the same running example in the E-Commerce
domain in order to illustrate the differences in the obtained results. Thereafter, we raise
the level of abstraction and we discuss the issues that we have to cope with when working
on the software side or the performance side to detect and remove performance problems.

The paper is organized as follows:
in Section \ref{sec:experiment} a running example in the E-Commerce domain
is presented and performance analysis results obtained by applying the two approaches
are illustrated; Section \ref{sec:discussion} discusses the major issues related to working
on the software or the performance side in a round-trip SPE process; Section \ref{sec:related}
presents related works, and finally Section \ref{sec:conclusions} concludes the paper. 
\section{Running Experiment}\label{sec:experiment}

In this section we show an application of the two approaches of interest in the
E-Commerce domain.
We first describe the E-Commerce System software architectural model, then numerical
results obtained from the experimentation are presented and discussed.

\subsection{The E-Commerce System (ECS)}\label{sec:ecs}

\subsubsection{The ECS model}\label{sec:ecs-model}

ECS is a web-based system that manages business data. We assume to have a multi-view model,
composed by (i) \emph{Static} and (ii) \emph{Dynamic} \emph{View}s.

\begin{figure}[!h]
   \centering
   \vspace{-0.3cm}
   \includegraphics[width=12cm]{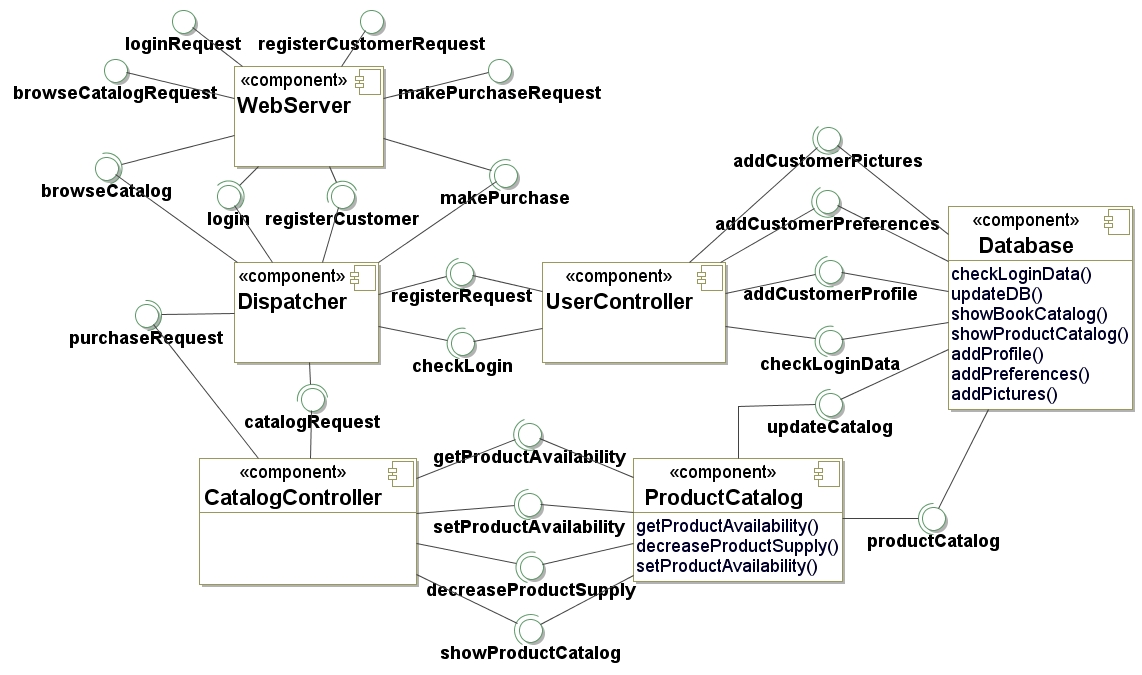}
   \vspace{-0.3cm}
   \caption{Initial ECS - Static View.}
   \label{fig:ecs-static}
   \vspace{-.7cm}
\end{figure}

\begin{figure}[!h]
 \centering
 \vspace{-0.3cm}
 \begin{tabular}{c}
 \subfigure[\emph{Register} service]
   {\includegraphics[width=12cm]{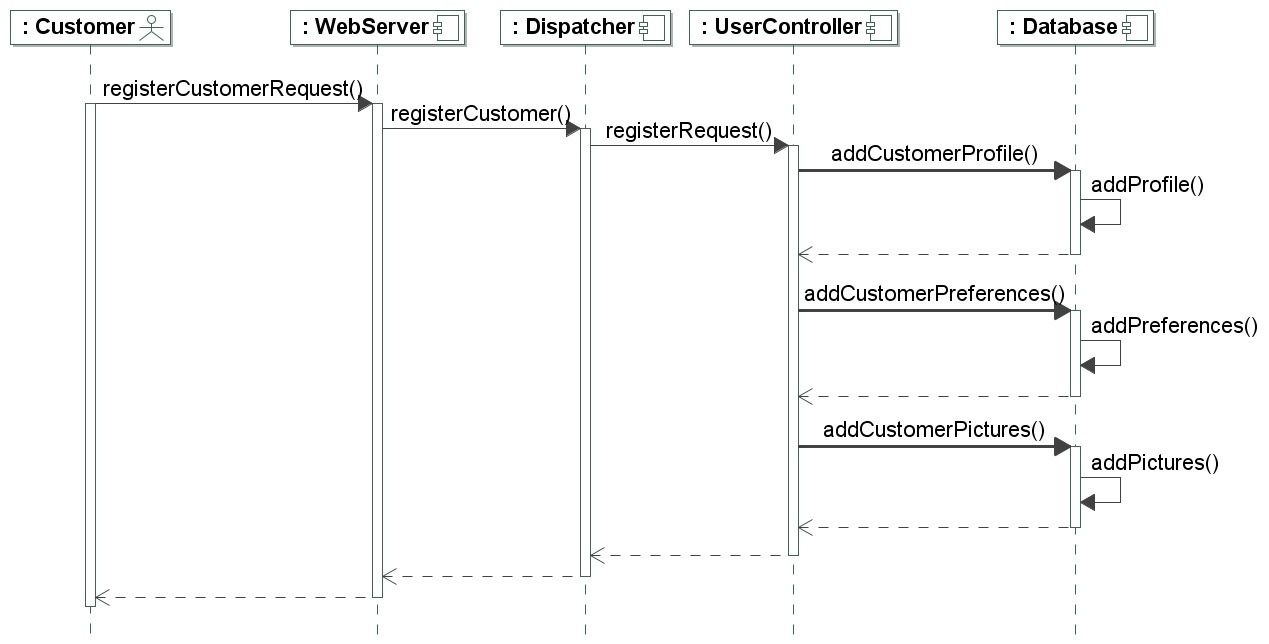}}\\
 \subfigure[\emph{BrowseCatalog} service]
   {\includegraphics[width=13cm]{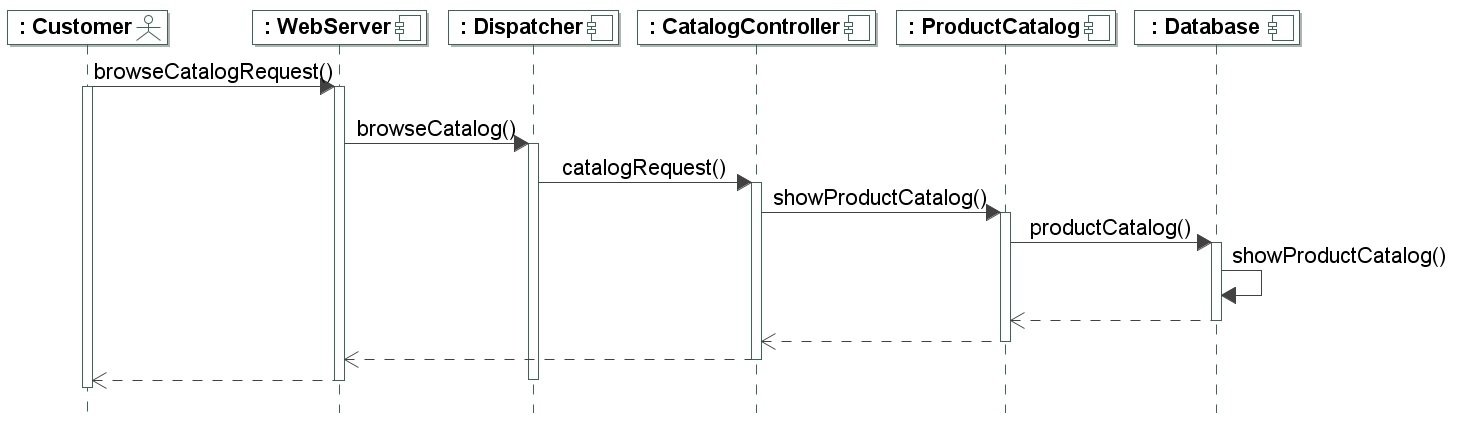}}\\
 \subfigure[\emph{MakePurchase} service]
   {\includegraphics[width=13cm]{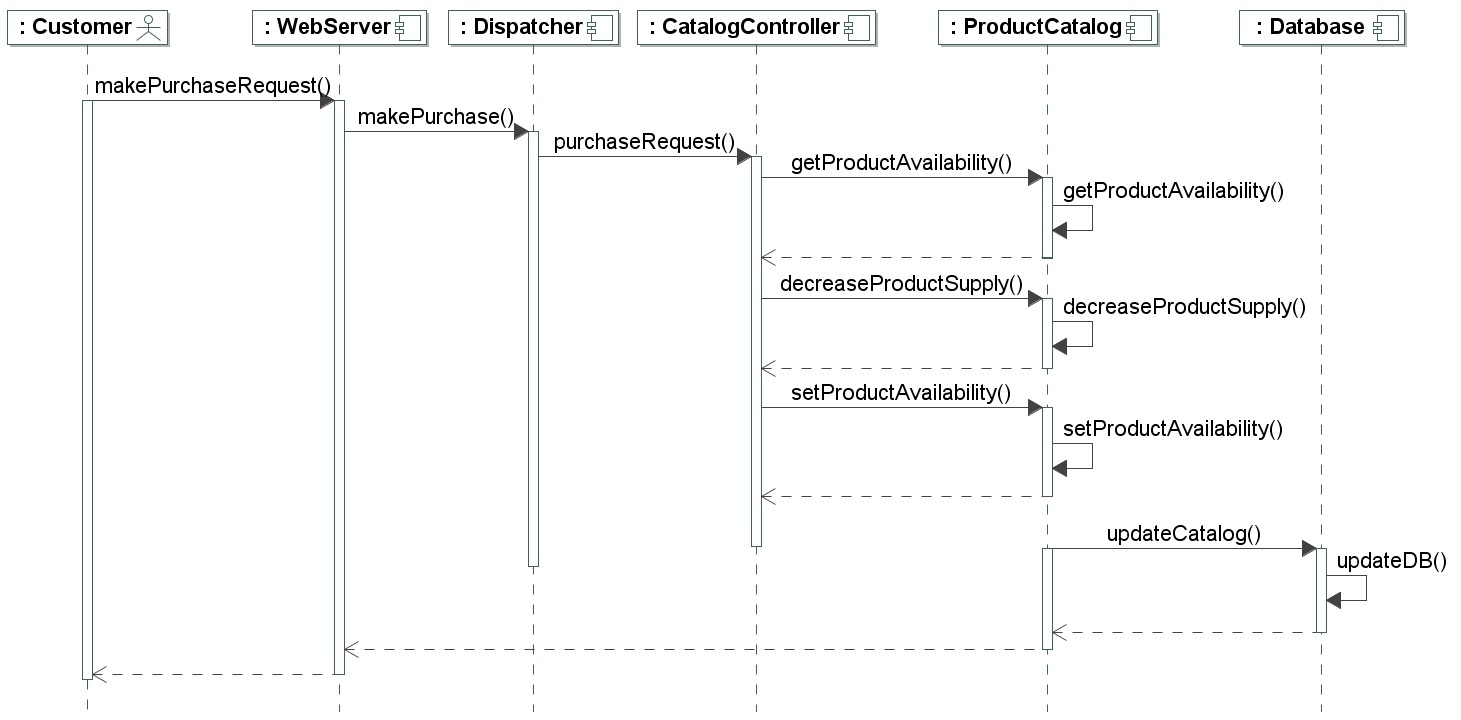}}
 \end{tabular}
 \vspace{-0.3cm}
 \caption{Initial ECS - Dynamic View.}
 \label{fig:ecs-dynamic}
 \vspace{-0.5cm}
\end{figure}

Several components are in the (server-side) \emph{Static View} of Figure \ref{fig:ecs-static},
each one providing/requiring interfaces and/or operations called during services execution.
Among all provided services in this paper we focus on the three ones of Figure
\ref{fig:ecs-dynamic}, namely: (a) \emph{Register} related to customers registration,
(b) \emph{BrowseCatalog} for consulting the products catalog, and (c) \emph{MakePurchase}
that is executed whenever a customer wants to purchase a product.

Note that, since we adopted the SAP\ensuremath{\bullet}one methodology \cite{SAPONE} for generating
a QN performance model \cite{primaUML} from a (platform-independent) UML software model in both
the compared approaches, no information about the deployment of software components is provided.
Hence, we are assuming that every software component is placed on a logical device,
and that all the devices have the same ``speed'' but they can manage requests through queues
of different capacity and different scheduling policy.
When the performance model is generated from the software model, that assumption allows us to directly
map each software component to a single service center, and to connect it with the other ones in respect
with the interface realizations/usages in the \emph{Static View} and the message flows in the
\emph{Dynamic View}.

%

%

\subsubsection{Performance requirements, legacy constraints and performance annotations for ECS}
\label{sec:ecs-requirements}

Several performance requirements have been defined on services response time and hardware devices
utilization:

\begin{itemize}
    \vspace{-.1cm}
    \item [\emph{R1:}] no service must have a response time greater than 4 seconds
    in the server-side when in the whole system there are 150, 300, and 50 users
    requesting respectively the \emph{MakePurchase}, \emph{BrowseCatalog}, and
    \emph{Register} services;
    \vspace{-.1cm}
    \item [\emph{R2:}] no hardware resource has to be used more than 90\% when in the
    whole system there are 150, 300, and 50 users requesting respectively the
    \emph{MakePurchase}, \emph{BrowseCatalog}, and \emph{Register} services.
    \vspace{-.1cm}
\end{itemize}
Furthermore, a legacy constraint has been introduced:
\begin{itemize}
    \vspace{-.1cm}
    \item [\emph{C1:}] since we assume that the \emph{Database} is a Commercial Off-The-Shelf
    component, it can neither be refactored in any way nor replaced by another database having
    better performance.
\end{itemize}
In order to carry out performance analysis, several input parameters for the
performance model obtained from the ECS model by means of the forward transformation have to
be defined:

\textbf{Workload characterization.}
Performance requirements define the average number of users for workloads. Hence, we define
a closed workload class for each considered service. In particular, the average number of
users for \emph{MakePurchase}, \emph{BrowseCatalog}, and \emph{Register} are, respectively,
150, 300, and 50. This means that under an aggregate average number of users equal to 500,
the 10\% is registering, the 60\% is browsing the products catalog, and the 30\% is
purchasing a product.

\begin{figure}[!h]
   \centering
   \vspace{-.2cm}
   \includegraphics[width=6.7cm]{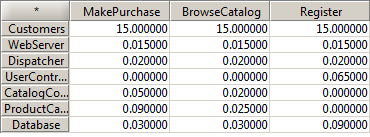}
   \vspace{-0.4cm}
   \caption{Service demands for the initial ECS.}
   \label{fig:ecs-demands}
   \vspace{-.6cm}
\end{figure}

\textbf{Service demands definition.}
Figure \ref{fig:ecs-demands} shows service demands of the considered services.
Values are expressed in seconds. Since \emph{Customer} generates workloads, we consider it
as the delay node whose think time is 15 seconds in order to represent the mean time needed by
a user for elaborating a request and the arrival of the latter to the server-side. All the other
service demands are orders of magnitude lower than think time because there is no time spent in
thinking. Note that service demands are well proportioned relating to requests in the various
services. Hence, we are confident about their adequacy.

\subsubsection{Performance analysis for ECS}
\label{sec:ecs-analysis}

All the performance analysis have been conducted by transforming each involved software architectural
model into a QN performance model \cite{primaUML}, and by solving the latter with the JMT tool
\cite{DBLP:conf/wosp/CasaleS11}.

\begin{figure}[!h]
 \centering
 \vspace{-.3cm}
 \begin{tabular}{c}
 \subfigure[Response times]
   {\includegraphics[width=6.2cm]{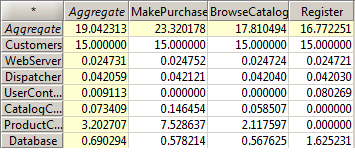}}
 \hspace{1cm}
 \subfigure[Utilizations]
   {\includegraphics[width=6.2cm]{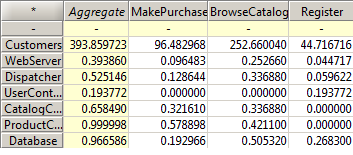}}
 \end{tabular}
 \vspace{-0.3cm}
 \caption{Response times and utilizations of the initial ECS.}
 \label{fig:ecs-results}
 \vspace{-0.3cm}
\end{figure}

Figure \ref{fig:ecs-results} shows performance analysis results for the QN corresponding to the
initial ECS model for the indices of interest, i.e. response times and utilizations.

As illustrated in Figure \ref{fig:ecs-results}.(a), \emph{R1} is widely
violated by the \emph{MakePurchase} service. In fact, under a workload of 150 users
purchasing a product, in the server-side the mean time elapsed from a single request
arrival to its departure (i.e. the response time) is 23.32 - 15.00 = 8.32 seconds,
i.e. more than double compared to the defined threshold of 4 seconds.
Instead, the response times at server-side for the other two services don't violate \emph{R1}.
In fact, \emph{BrowseCatalog} has a mean response time of
17.81 - 15.00 = 2.81 seconds whereas \emph{Register} has a mean response time of
16.77 - 15.00 = 1.77 seconds.

As illustrated in Figure \ref{fig:ecs-results}.(b), \emph{R2} is
violated by \emph{Database} and \emph{ProductCatalog}. In fact, under the specified
workload, the former has an utilization of 96.66\% whereas the latter has an
utilization of 99.99\%.

Since several performance indices of interest are not satisfactory we need to
refactor the software model in order to satisfy violated requirements.

\subsection{ECS refactoring based on performance antipatterns}
\label{sec:ecs-ref-ap}

In this section we perform two refactorings on the initial ECS model by using the
approach based on performance antipatterns \cite{QoSA2012}. We first execute the
antipatterns detection phase in order to identify bad practices related to performance,
then we randomly choose an antipattern to remove. Hence, given the corresponding
pair (problem, solution), we execute a refactoring that applies the solution to the
problem.

\begin{figure}[!h]
   \centering
   \vspace{-0.3cm}
   \includegraphics[width=12cm]{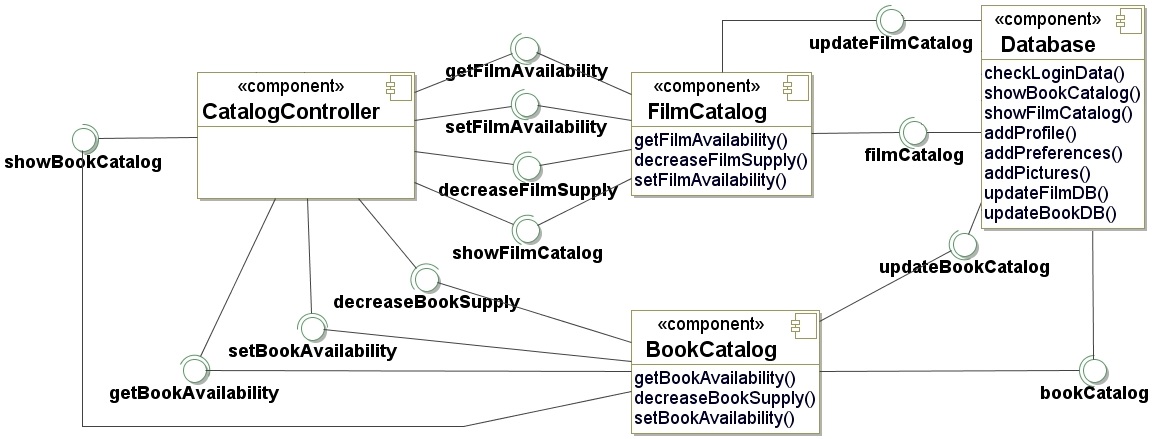}
   \vspace{-0.2cm}
   \caption{Refactored ECS (BLOB removed) - Static View.}
   \label{fig:ecs-ref1-ap-static}
   \vspace{-0.3cm}
\end{figure}

Let us assume that, among all the detected antipatterns, we choose to remove the
occurrence of the BLOB antipattern where the so called ``Blob''
entity is \emph{ProductCatalog}. As stated in \cite{DBLP:conf/cmg/SmithW03a},
this antipattern \emph{``occurs when a single class either 1) performs all of the
work of an application or 2) holds all of the application's data''} and it can be
solved \emph{``by refactoring the design to distribute intelligence uniformly over the
application's top-level classes, and to keep related data and behavior together''}.
Hence, let us assume that products managed by ECS are films and books and that these
types of products are requested respectively for 80\% and 20\%. We can split the
\emph{ProductCatalog} in two components, i.e. \emph{FilmCatalog} and \emph{BookCatalog}.
Figure \ref{fig:ecs-ref1-ap-static} shows the refactored ECS static view, where the
two components have been introduced and adequately connected to the
other ones. Note that their interfaces/operations are conceptually coherent with
data each one of them manages. Also dynamic and deployment views are affected by
the refactoring. In particular, in the dynamic view a probability-weighted alternative
fragment with two operands related to films (with a probability of 0.8) and books
(with a probability of 0.2) replaces each portion of \emph{MakePurchase} and
\emph{BrowseCatalog} services involving \emph{ProductCatalog}, replicating that
portion for both \emph{FilmCatalog} and \emph{BookCatalog} in an adequate manner
\footnote{Readers interested to all
details related to the refactored views not shown in this paper can refer to the external
resource available at \url{http://www.di.univaq.it/cortelle/docs/FESCA2013\_appendix.pdf}.}.

\begin{figure} [!h]
 \centering
 \vspace{-0.4cm}
 \begin{tabular}{c}
 \subfigure[Response times]
   {\includegraphics[width=6.2cm]{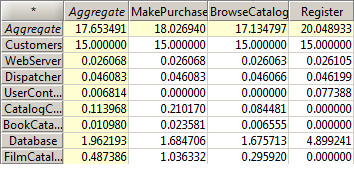}}
 \hspace{1cm}
 \subfigure[Utilizations]
   {\includegraphics[width=6.2cm]{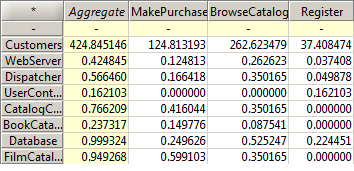}}
 \end{tabular}
 \vspace{-0.3cm}
 \caption{Response times and utilizations after the BLOB antipattern removal.}
 \label{fig:ecs-ref1-ap-results1}
 \vspace{-0.3cm}
\end{figure}

Figure \ref{fig:ecs-ref1-ap-results1} shows performance analysis results of the QN
corresponding to the ECS model for the refactoring described in this section.

As illustrated in Figure \ref{fig:ecs-ref1-ap-results1}.(a), \emph{R1} is
no longer violated for the \emph{MakePurchase} service. In fact, under a workload of
150 users purchasing a product, the response time at the server-side is 18.03 - 15.00
= 3.03 seconds, that are almost one second lower than the defined threshold of 4 seconds.
Also the response time at server-side for \emph{BrowseCatalog} improves, hence it still
not violate \emph{R1}. Instead, the response times at server-side for
\emph{Register} significantly increases, and it becomes 20.05 - 15.00 = 5.05 seconds,
i.e. greater than the defined threshold of 4 seconds.

As illustrated in Figure \ref{fig:ecs-ref1-ap-results1}.(b), \emph{R2} is
violated for \emph{Database} and \emph{FilmCatalog}.
In fact, under the specified workload, the former has an utilization of 99.9\% whereas
the latter has an utilization of 95\%.

Since several performance indices of interest are not satisfactory we need
further refactoring in order to satisfy violated requirements.



Let us assume that, among all the detected antipatterns, we choose to remove the
occurrence of the EST antipattern in \emph{Register}, in order to obtain a better
response time for that service\footnote{Note that the EST occurrence detected in
this second step could also have been detected in the previous iteration where we
opted for the BLOB occurrence removal.}.
As stated in \cite{DBLP:conf/cmg/SmithW03a}, this antipattern
\emph{``occurs when an excessive number of requests is required to perform a task.
It may be due to inefficient use of available bandwidth, an inefficient interface,
or both''} and it can be solved by means of the application of the Session Facade
design pattern \cite{javaSun2001}.
Hence, two new communicating components, i.e. \emph{RemoteFacade} and
\emph{LocalFacade}, are introduced between the one originating an excessive number
of requests, i.e. \emph{UserController}, and the destination of those requests,
i.e. \emph{Database}.
This refactoring results in
more efficient interfaces and also in a more efficient use of available bandwidth.

\begin{figure}[!h]
   \centering
   \vspace{-.3cm}
   \includegraphics[width=12cm]{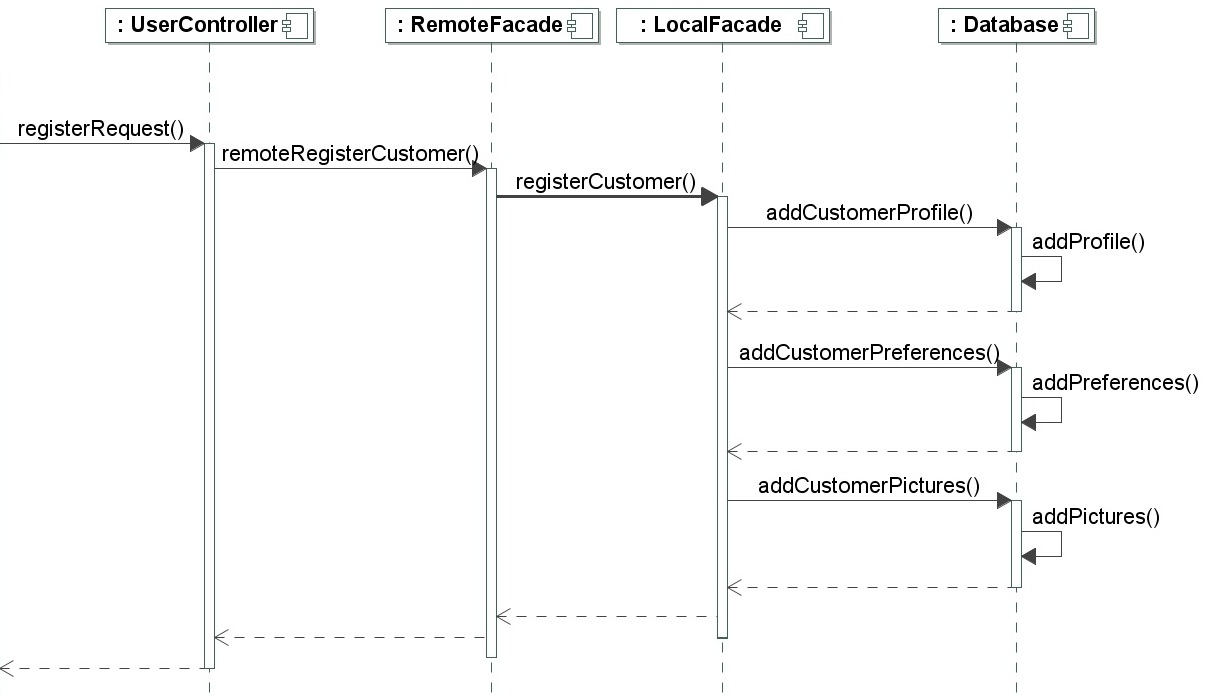}
   \vspace{-.3cm}
   \caption{Refactored ECS (EST removed) - Dynamic View (\emph{Register} service)}
   \label{fig:ecs-ref2-ap-dynamic}
   \vspace{-.3cm}
\end{figure}

Figure \ref{fig:ecs-ref2-ap-dynamic} shows the refactored \emph{Register} service of
ECS, where the two Facade components have been adequately introduced in the
dynamics of that service. Also static and deployment views are affected by
the refactoring. In particular, also in the static view the two Facade components and
their interfaces have been adequately introduced.

\begin{figure}[!h]
 \centering
 \begin{tabular}{c}
 \subfigure[Response times]
   {\includegraphics[width=6.2cm]{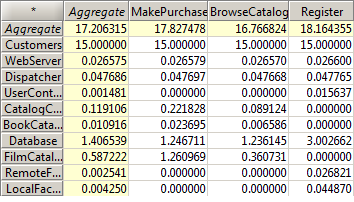}}
 \hspace{1cm}
 \subfigure[Utilizations]
   {\includegraphics[width=6.2cm]{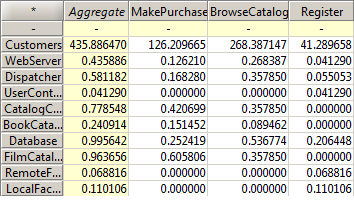}}
 \end{tabular}
 \vspace{-.3cm}
 \caption{Response times and utilizations after the EST antipattern removal.}
 \vspace{-0.5cm}
 \label{fig:ecs-ref2-ap-results2}
\end{figure}

Figure \ref{fig:ecs-ref2-ap-results2} shows performance analysis results of the QN
corresponding to the ECS model for the refactoring described in this section.

As illustrated in Figure \ref{fig:ecs-ref2-ap-results2}.(a), \emph{R1} is
no longer violated for the \emph{Register} service. In fact, under a workload of
50 users requesting a registration, the response time at the server-side is 18.16 - 15.00
= 3.16 seconds, that are almost one second lower than the defined threshold of 4 seconds.
Also response times at server-side for \emph{BrowseCatalog} and \emph{MakePurchase} improve,
hence they still not violate \emph{R1}. In fact, the response time at server-side
for the former becomes 16.77 - 15.00 = 1.77 seconds, whereas the one for the latter
becomes 17.83 - 15.00 = 2.83 seconds.

As illustrated in Figure \ref{fig:ecs-ref2-ap-results2}.(b), \emph{R2} is
violated for \emph{Database} and \emph{FilmCatalog}.
In fact, under the specified workload, the former has an utilization of 99.57\% whereas
the latter has an utilization of 96.37\%. Finally, utilizations of \emph{RemoteFacade}
and \emph{LocalFacade} are, respectively, 6.89\% and 11.01\%, hence they do not
violate \emph{R2}.

Since several performance indices of interest are not satisfactory we need further
refactoring in order to satisfy violated requirements but, at this point, we stop iterating
the antipattern-based refactoring approach and turn back to the initial ECS model in order
to start the application of the refactoring approach that uses bidirectional transformations.

\subsection{ECS refactoring based on bidirectional transformations}\label{sec:ecs-ref-bt}

In this section we perform two refactorings on the initial ECS model by using the
approach based on bidirectional transformations. Hence, refactorings are executed by the
performance analyst directly on the performance model, basing on his own expertise on
interpreting performance results.

\begin{figure}[!h]
   \centering
   \vspace{-0.3cm}
   \includegraphics[width=12cm]{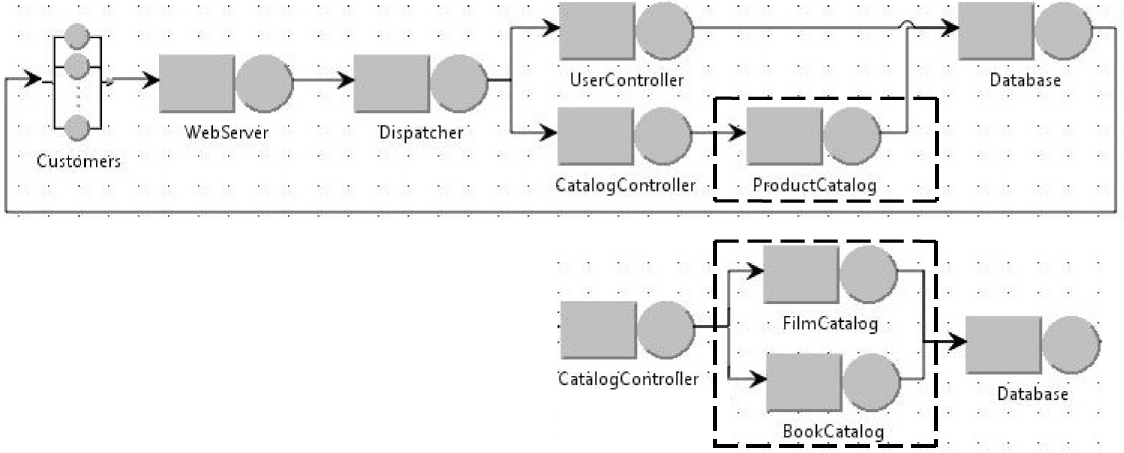}
   \vspace{-0.3cm}
   \caption{QN refactoring for ECS (\emph{ProductCatalog} split)}
   \label{fig:ecs-ref1-bt-qnm}
   \vspace{-0.3cm}
\end{figure}

Starting from the QN model in the top-side of Figure \ref{fig:ecs-ref1-bt-qnm} 
resulting from the application of the forward transformation to the initial ECS model, 
performance results of Figure \ref{fig:ecs-results} are obtained.
As we stated in Section \ref{sec:ecs-ref-ap}, we assume that products managed by ECS are films
and books and that these types of products are requested respectively for 80\% and 20\%.
Since the \emph{Database} cannot be refactored in any way, we assume that the performance
analyst refactors the QN model as shown in dashed boxes of Figure \ref{fig:ecs-ref1-bt-qnm},
hence by splitting \emph{ProductCatalog} (i.e. the bottleneck) in two service centers, i.e.
\emph{FilmCatalog} and \emph{BookCatalog}, for managing the two types of products.
Service demands for new service centers are adjusted basing on percentiles assumed above.

Note that, by applying the backward transformation in order to go back to the software model,
the same refactored ECS resulting from the first refactoring of Section \ref{sec:ecs-ref-ap}
is obtained. Therefore, also the performance results are the same, hence several performance
indices of interest are not satisfactory and we need further refactoring at  performance model
side in order to satisfy violated requirements.

\begin{figure}[!h]
   \centering
   \vspace{-0.6cm}
   \includegraphics[width=12cm]{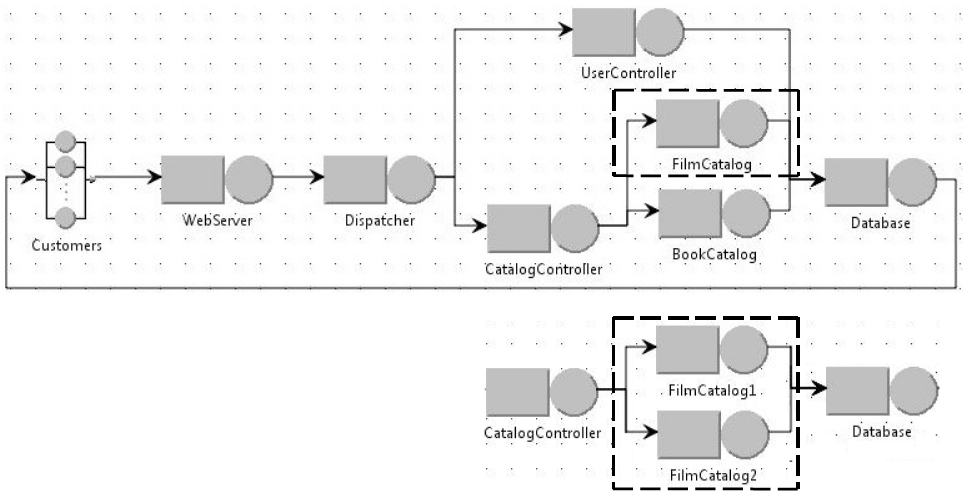}
   \vspace{-0.3cm}
   \caption{QN refactoring for ECS (\emph{FilmCatalog} split).}
   \label{fig:ecs-ref2-bt-qnm}
   \vspace{-0.6cm}
\end{figure}

Let us assume that, since the \emph{Database} cannot be refactored in any way,
the performance analyst refactors the QN model as shown in dashed boxes of Figure
\ref{fig:ecs-ref2-bt-qnm}, hence by splitting \emph{FilmCatalog}
(i.e. the bottleneck) in two service centers, i.e. \emph{FilmCatalog1} and
\emph{FilmCatalog2}, and balancing their load. This means that requests for films are
equally distributed between the two service centers, i.e. \emph{CatalogController}
routes the requests to the two service centers with an identical probability of 0.5.
Service demands for new service centers are adjusted basing on that probability.

\begin{figure}[!h]
   \centering
   \vspace{-0.3cm}
   \includegraphics[width=11cm]{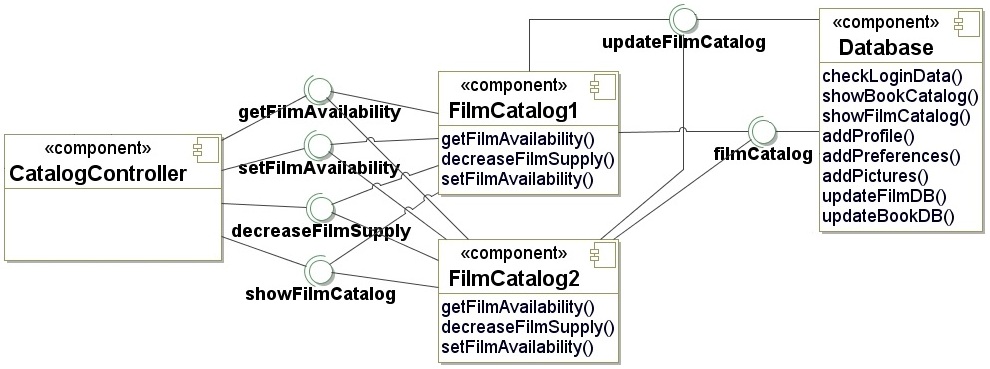}
   \vspace{-0.3cm}
   \caption{Refactored ECS resulting from QN refactoring (\emph{FilmCatalog} split) - Static View.}
   \label{fig:ecs-ref2-bt-static}
   \vspace{-0.3cm}
\end{figure}

Figure \ref{fig:ecs-ref2-bt-static} shows an excerpt of the static view for the
refactored ECS obtained applying the backward transformation in order to go back to the
software model. Also dynamic and deployment views are affected by the refactoring.
In particular, in the dynamic view a probability-weighted alternative fragment with two
operands related to request balancing (probability of 0.5) replaces the portion of
\emph{MakePurchase} and \emph{BrowseCatalog} services involving \emph{FilmCatalog},
by replicating that portion for both \emph{FilmCatalog1} and \emph{FilmCatalog2}
in an adequate manner.

\begin{figure}[!h]
 \centering
 \begin{tabular}{c}
 \subfigure[Response times]
   {\includegraphics[width=6.2cm]{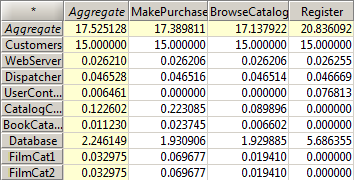}}
 \hspace{1cm}
 \subfigure[Utilizations]
   {\includegraphics[width=6.2cm]{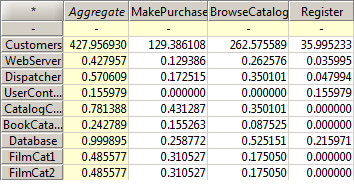}}
 \end{tabular}
 \vspace{-0.4cm}
 \caption{Response times and utilizations after the \emph{FilmCatalog} split.}
 \label{fig:ecs-ref2-bt-results2}
 \vspace{-0.5cm}
\end{figure}

Figure \ref{fig:ecs-ref2-bt-results2} shows performance analysis results of the QN
corresponding to the ECS model for the refactoring described in this section.

As illustrated in Figure \ref{fig:ecs-ref2-bt-results2}.(a), \emph{R1} is
still not violated for \emph{MakePurchase} and \emph{BrowseCatalog} services.
In fact, under a workload of 150 users purchasing a product, the response time at
the server-side for the former is 17.39 - 15.00 = 2.39 seconds, resulting in an
improvement greater than half a second whereas, under a workload of 300 users
browsing the catalog, the response time at the server-side for the latter remains
more or less unchanged, i.e. 17.14 - 15.00 = 2.14 seconds (improvement of 0.25 seconds).
Instead, the response time at server-side for \emph{Register} increases, and it
becomes 20.85 - 15.00 = 5.85 seconds, i.e. still greater than the defined threshold
of 4 seconds.

As illustrated in Figure \ref{fig:ecs-ref2-bt-results2}.(b), the only component
having an utilization that violates \emph{R2} is \emph{Database}.
In fact, under the specified workload, it has an utilization of 99.9\%.

Since several performance indices of interest are not satisfactory we need further
refactoring in order to satisfy violated requirements but, at this point, we stop
iterating the refactoring approach that uses bidirectional transformations in order to
compare both the approaches on the same number of iterations.
%
\section{Raising the abstraction level}\label{sec:discussion}

In this section, in light of the example illustrated in Section \ref{sec:experiment},
we discuss the issues that characterize refactoring approaches that work on the software
side and the ones working on the performance side.

\subsection{Level of automation and human role}

In order to be adopted in practice, a round-trip SPE process needs to be supported by
a high level of automation, due to the complexity of tasks that have to be executed
and the decisions that have to be taken. Nevertheless, the human role in SPE should not
be completely removed, because the experience and skills of software designers and/or
performance analysts cannot be fully embedded in automated processes, as we discuss
in the following.

Let us look at the steps in the forward path of Figure \ref{fig:process}: software to performance
model transformations have been fully automated in the last decade even for high complexity
cases \cite{DBLP:books/daglib/0027475}; performance model solution is a well assessed task since
decades and very sophisticated techniques are available today \cite{DBLP:conf/wosp/CasaleS11}.

So, let us focus on the backward path of Figure \ref{fig:process}. 
Automation in the problem detection and solution step is traditionally very well supported on the
performance side, where a whole theory on bottleneck identification and removal has been introduced
few decades ago (e.g. \cite{Lazowska:1984:QSP:2971}) and has been continuously refined by
more recent results (e.g. \cite{bottleneckWood}). However, refactoring actions applied on the performance 
side have to be reported on the software side.
The automation of the former task rests on bidirectional model transformations, as shown in Section
\ref{sec:ecs-ref-bt}. Bidirectional transformations are complex to build in this domain,
mostly due to the low injectivity of forward transformations \cite{Sabetta2005} that usually collapse
several elements of a software model into an unique element of a performance model, making back-tracing
more difficult. Instead, on the software side a certain level of automation has been introduced only
recently, for example based on antipatterns (as illustrated in our example) or on metaheuristics that
search the solution space looking for changes that can improve the performance indices \cite{Martens2010}.

However, in both cases it may be necessary to decide among alternative refactoring actions,
because it is difficult to implement such a sharp detection and solution step that terminates
with an unique suggestion of refactoring actions, especially in large scale systems.

The number of alternative refactoring actions outcoming from the detection and solution
phase on the software side can be considerably higher than the one on the performance side.
This is mostly due from the richness of software model notations that makes the space
of possible solutions quite larger than the one on the performance side. For example, the reduction
of number of visits to a network node in a QN performance model consists in modifying an
integer parameter, whereas in UML may correspond to several alternatives that reduce the
number of messages exchanged in a behavioral model (i.e. the EST removal in Section \ref{sec:ecs-ref-ap}).
This aspect, on one end, can be considered as an advantage of approaches that work on the software side,
because they propose a variety of choices to software designers that can select the most appropriate one(s),
for example considering cost factors or legacy constraints.
On the other end, a too large variety of solutions (i.e. the list of detected antipatterns) can be hard
to manage. Decision support mechanisms, for example based on convenience metrics, might help the designer
to mitigate this aspect \cite{APguiltiness2010, Martens2010}. However, they are heuristic techniques that
sometimes do not work better than the designer's experience.

\subsection{Effectiveness of refactoring actions}

Independently of the considered side, after a set of alternative refactoring actions will be determined,
either the software designer or the performance analyst has to decide which one(s) applying to the
(software or performance) model. This is a key decision for sake of process convergence. In fact, a
decision tree drives the process, where each branch (i.e. each set of actions applied) leads to a
new model that has to be solved in order to check whether performance problems have been removed
(e.g. a performance requirement, that was violated, is now satisfied).

As mentioned above, heuristic techniques (possibly based on convenience metrics) might support this decision
both on the software and the performance side. However, the effectiveness of refactoring actions will be
given by the tradeoff between the refactoring complexity (i.e. the distance between the original model
and the refactored one) and the performance gain obtained from the refactoring. For example, it may happen
that heavy re-design actions, like splitting software components and modifying their interaction patterns
(i.e. \emph{FilmCatalog} and \emph{BookCatalog} that replace \emph{ProductCatalog} in Section
\ref{sec:ecs-ref-ap}), bring little performance benefits as compared to re-deploy a software component
(\emph{ProductCatalog}) on another processing node (i.e. \emph{DatabaseNode} of ECS).
In most cases it is difficult to predict the performance gain of a refactoring action without actually
solving a refactored performance model.
However, again due to the usual richness of software modeling notations, the refactoring complexity is
generally lower on the performance side. At least, the portfolio of refactoring actions that can be applied,
for example on a Queueing Network, is certainly more limited than the one working on an UML model.
On one end, this implies that in order to solve nested performance problems more iterations may be needed
on the performance side compared to the software side. On the other end, the performance gain of refactoring
actions applied to the software side is more unpredictable due to the forward transformation that generates
a refactored performance model from a refactored software model. In worst cases, software refactoring actions
could lead to performance degradation due to side effects that are not visible in the software model (e.g.
false positive performance antipatterns)\footnote{This problem is analogous to looking for bugs in software
code: working on high-level languages (like C) gives a quite rich syntax, but modifications must be translated
in assembler before looking at their effects, whereas working in assembler provides a more direct control
on the effects of changes.}.

\subsection{Scalability}

Several factors contribute to the scalability of refactoring approaches. 
First, it comes straightforward that a single loop of the round-trip process illustrated in Figure \ref{fig:process}
is shorter in case of refactoring on the performance side than on the software side. This is due to the need of
applying a forward transformation to a refactored software model for obtaining a refactored performance model.
Direct refactoring on the performance side, however, requires at the end of the process (i.e. when a satisfactory
performance model has been obtained) the application of a backward transformation to build back a corresponding 
satisfactory software model. 
This observation leads to the scalability of the transformations. 

Let us assume that the forward transformation has a
$O(forward)$ complexity, whereas the complexity of the forth and back transformations are,
respectively, $O_{bid}(forth)$ and $O_{bid}(back)$\footnote{Note that the complexity of the forward transformation
in the context of a bidirectional one is higher than a simple forward transformation, because tracing information
has to be brought to obtain the backward transformation \cite{DBLP:conf/qosa/EramoCPT12}.}. By a rough counting, if N
iterations are necessary to solve problems on the performance side and M iterations (usually with $M<N$) are
necessary on the software side, then the scalability tradeoff can be expressed by:

\vspace{-.5cm}
$$M*O(forward)<N*(O_{bid}(forth)+O_{bid}(back))$$
\vspace{-.7cm}

Of course, the transformation complexities depend on the distance between the specific source software metamodel and target
performance metamodel.

Another factor that affects the scalability is the number of iterations necessary to obtain satisfying results. This
translates to the depth of the decision tree that we have mentioned above. As discussed in the previous subsection,
this obviously depends on the effectiveness of refactoring actions, but also on their dependencies. In fact, alternative
refactoring actions that can be independently applied (i.e. the refactored model obtained by applying all of them does not
depend on the application order, as the sequential removal of BLOB and EST antipatterns of Section \ref{sec:ecs-ref-ap}),
ease the tree navigation. As opposite, alternative actions that affect each other, as splitting a component introduced
in a previous refactoring (i.e. the \emph{FilmCatalog} splitting of Section \ref{sec:ecs-ref-bt}), requires
a (possibly expensive) backtrack process on the decision tree to be considered.
Heuristic approaches to prune the decision tree would be suitable in this context. However, it looks easier (but likely
less effective) to decide on the performance side than on the software side. For example, it is usually more evident
the bottleneck to be first removed than the antipattern to be solved (among alternative ones).

\section{Related work}\label{sec:related}

At best of our knowledge this is the first paper that compares approaches for
model refactoring based on performance analysis. Hence, we present a brief
overview of works related to the approaches we compared, i.e. \cite{QoSA2012} and
\cite{DBLP:conf/qosa/EramoCPT12}.

\textbf{Working on the software model side.}
Very few model-based approaches for automated performance diagnosis and improvement
have been introduced up today in the software modeling domain.

In \cite{Martens2010} and \cite{DBLP:conf/qosa/KoziolekKR11}
meta-heuristic search techniques are used for improving different non-functional
properties of component based software systems by means of evolutionary algorithms.
The main limitation of such approaches is that it is quite time-consuming because
the design space may be huge.

In general, there has been a significant effort in software  refactoring based on
design patterns \cite{DBLP:journals/tse/FranceKGS04}. However, differently from patterns,
antipatterns look at the negative features of a software system and describe commonly
occurring solutions to problems that generate negative consequences \cite{Brown, Laplante}.

\emph{Technology-independent} performance antipatterns have been defined in
\cite{DBLP:conf/cmg/SmithW03a} and they represent the main references in our AP-based
works \cite{QoSA2012, DBLP:conf/iceccs/CortellessaMT10}. \emph{Technology-specific}
antipatterns have been specified in \cite{Dudney2003, Tate2003} and, more recently, in
\cite{DBLP:journals/jot/ParsonsM08}, where EJB antipatterns have been represented as a set
of rules loaded into a detection engine and they are detected by a monitoring mechanism
that leads to get run-time system properties which are matched with pre-defined rules.

\textbf{Working on the performance model side.}
The formal definition of a round-trip engineering process considering the
non-totality and non-injectivity of model transformations is presented in
\cite{Hettel2008}. Valid modifications on target models are limited to the ones
which do not induce backward mappings out the source metamodel and are not operated
outside the transformation domain.

In \cite{DBLP:conf/wosp/Xu08} performance problems are identified through the detection of bottlenecks
and long paths on Layered Queueing Networks (LQN) models. Contrary to our approach based
on bidirectional transformations \cite{DBLP:conf/qosa/EramoCPT12}, however, in \cite{DBLP:conf/wosp/Xu08}
no clue is given on the back propagation of performance model changes to any software model
notation.

Stevens \cite{DBLP:conf/models/Stevens07} discusses bidirectional transformations
focusing on basic properties which they should satisfy and pointing out some ambiguity
about specification of non-bijective transformations.

Bidirectional transformations based on triple graph grammars (TGGs) are presented in
\cite{Konigs2006}, where models are interpreted as graphs and transformations are
executed by using graph rewriting techniques.

Recently some interesting solutions based on lenses have been proposed.
In \cite{DBLP:journals/jot/DiskinXC11} the authors illustrate a technique to support
bidirectional transformations relying no more on mapping between models but on
differences operable on them. 
\section{Conclusions}
\label{sec:conclusions}

In this paper we have compared two performance-based model refactoring approaches
at work on the same example. This allowed us to highlight the peculiar aspects of
working on either the software side or the performance side. We have finally 
summarized our findings in Table \ref{tb:comparison}.

   \begin{table}[h]
   \vspace{-.2cm}
   \scriptsize
   \begin{center}
   \begin{tabular}{|c|c|c|c|}
           \hline
           \multicolumn{2}{|c|}{Classification parameters} & AP-based approach (software side) & BT-based approach (performance side)\\
           \hline
           \hline
           \multirow{2}{*}{\begin{minipage}{0.8in}\centering Human skills required\end{minipage}} & Design skills & medium & low\\
           \cline{2-4}
           & Performance skills & low & high\\
           \hline
           \multirow{2}{*}{\begin{minipage}{0.5in}\centering Degree of automation\end{minipage}} & Problem detection & medium & high\\
           \cline{2-4}
           & Problem solution & medium & high\\
           \hline
           \multirow{3}{*}{\begin{minipage}{0.5in}\centering Refactoring metrics\end{minipage}} & Number of actions & high & low\\
           \cline{2-4}
           & Complexity & low, medium, high & low\\
           \cline{2-4}
           & Performance gain predictability & low & medium\\
           \hline
           \multirow{2}{*}{\begin{minipage}{0.5in}\centering Scalability\end{minipage}} & Number of iterations & low, medium & medium, high\\
           \cline{2-4}
           & Single iteration complexity & $O(forward)$ & $O_{bid}(forth)+O_{bid}(back)$\\
           \hline
   \end{tabular}
   \end{center}
   \vspace{-.4cm}
   \caption{Overview of the compared approaches.}
   \vspace{-0.3cm}
   \label{tb:comparison}
   \end{table}

With the increasing interest on this type of refactoring approaches we
retain very relevant to study contexts where different techniques can
better work than other ones. 
This is an initial step for the comparison of such approaches.
The afterthoughts of this experience will be either consolidated or turned down
by applying the approaches to a significant amount of examples, that is our
intent in the near future.

Several future directions will be investigated:
(i) it could be very interesting to study a mixed approach that combines
    the ones we compared in this paper, i.e. by executing sequences of
    bottleneck analysis followed by detection and solution of performance
    antipatterns, and/or viceversa;
(ii) the introduction of performance antipatterns at the performance model
    side could support the performance analyst in detection and solution of
    performance problems, although it should be considered that the number of
    alternative actions resulting from the detection and resolution phases on a
    performance model can be considerably fewer than the ones on the software side,
    due to the lower number of elements in performance models as compared to
    software models;
(iii) finally, we are working on the introduction of measurement-based
    performance problem detection and solution at the code level by means of
    monitoring-driven testing techniques for cloud applications\cite{CloudScale}.

\section{Acknowledgments}
This work has been supported by the European Office
of Aerospace Research and Development (EOARD), Grant/Cooperative
Agreement Award no. FA8655-11-1-3055 on ``Consistent evolution of
software artifacts and non-func\-tio\-nal models''.

\section{Bibliography}
\renewcommand\refname{}
\vspace{-1cm}

\bibliographystyle{eptcs}
\bibliography{biblio}


\end{document}